\documentclass[acmsmall, nonacm=true]{acmart}

\usepackage{gensymb}

\AtBeginDocument{%
  \providecommand\BibTeX{{%
    \normalfont B\kern-0.5em{\scshape i\kern-0.25em b}\kern-0.8em\TeX}}}

\setcopyright{acmcopyright}
\copyrightyear{2021}
\acmYear{2021}

\begin{document}
\title{Energy and Thermal-aware Resource Management of Cloud Data Centres:  A Taxonomy and Future Directions}
\author{Shashikant Ilager}
\orcid{0000-0003-1178-6582}

\author{Rajkumar Buyya}
\affiliation{%
  \institution{\\University of Melbourne}
  \department{Cloud Computing and Distributed Systems (CLOUDS) Laboratory, School of Computing and Information Systems}
  \streetaddress{Parkville Campus}
  \city{Melbourne}
  \state{Victoria}
  \postcode{3010}
  \country{Australia}}
\email{shashi.ilager@unimelb.edu.au}

\renewcommand{\shortauthors}{S. Ilager and R. Buyya}
\begin{abstract}
This paper investigates the existing resource management approaches in Cloud Data Centres for energy and thermal efficiency. It identifies the need for integrated computing and cooling systems management and learning-based solutions in resource management systems. A taxonomy on energy and thermal efficient resource management in data centres is proposed based on an in-depth analysis of the literature. Furthermore, a detailed survey on  existing approaches is conducted according to the taxonomy and  recent advancements including machine learning-based resource management approaches and cooling management technologies are discussed. 
Finally, key future research directions for sustainable growth of Cloud computing services are  proposed.
\end{abstract}

\keywords{energy efficiency, cloud computing, IoT, machine learning}

\maketitle

\section{Introduction}

Internet-based Distributed Computing Systems (DCS) such as Clouds have become an essential backbone of the modern digital economy, society, and industrial operations. The emergence of the Internet of Things (IoT), diverse mobile applications, smart grids, smart industries, and smart cities has resulted in massive amounts of data generation. Thus, increasing the demand for computing resources \cite{IoT} to process this data and derive valuable insights for users and businesses. According to the report from Norton \cite{norton2019usa},  21 billion IoT devices will be connected to the internet by 2025, creating substantial economic opportunities. 

\par Computing models such as Cloud have revolutionised the way services are delivered and consumed by providing flexible on-demand access to services with a pay-as-you-go model. Besides, new application and execution models like micro-services and serverless or Function as Service (FaaS) computing \cite{baldini2017serverless} are becoming mainstream that significantly reduces the complexities in the design and deployment of software components.
On the other hand, this increased connectivity and heterogeneous workloads demand distinct Quality of Service (QoS) levels to satisfy their application requirements \cite{gan2019open}\cite{dastjerdi2016fog}\cite{fox2009above}. These developments have led to the building of hyper-scale data centres and complex multi-tier computing infrastructures.

The Cloud data centres are the backbone infrastructures of Cloud computing today. A data centre is a complex Cyber-Physical-System (CPS) consisting of numerous elements. It houses thousands of rack-mounted physical servers, networking equipment, sensors (monitoring server and room temperature), a cooling system to maintain acceptable room temperature, and many other facility-related subsystems. It is one of the highest power density CPS, consuming up to 30-40 kW per rack, dissipating an enormous amount of heat. This poses a severe challenge to manage resources energy efficiently and provide reliable services to users. Moreover, even a 1\% improvement in data centre efficiency leads to savings in millions of dollars over a year and reduces the carbon footprints \cite{torell2015unexpected}.

\par Resource management in data centres is extremely challenging due to complex subsystems and heterogeneous workload characteristics.  It is impossible to fine-tune the controllable parameters by resource management systems manually. For example, \textit{``Just $10$ pieces of equipment, each with $10$ settings, would have $10$ to the $10^{th}$ power, or $10$ billion, possible configurations — a set of possibilities far beyond the ability of anyone to test for real''} \cite{schwartz2019allen}\cite{amodei2018ai}. Moreover, these large-scale systems have numerous subsystems interacting with each other and often have a non-linear relationship between their parameters. However, optimising data centre operation requires tuning the hundreds of parameters belonging to different subsystems where heuristics or static solutions fail to yield a better result.  Therefore, optimising these data centres using suitable Artificial Intelligence (AI) techniques is of great importance.

\par There have been many efforts in this regard using ML for systems focusing on optimising different computing layers \cite{JeffMLforSystem}. Public Cloud service providers and the data centre industry have also explored energy and thermal efficient resource management solutions using ML techniques.  For instance, ML-centric Cloud \cite{bianchini2020toward} is an ML-based RMS system at an inception stage from the Microsoft Azure Cloud. They built Resource Control (RC)—a general ML and prediction serving system that provides insights into the Azure compute fabric resource manager's workload and infrastructure. Similarly, Google has also applied ML techniques to optimise the efficiency of their data centres. Specifically, they have used ML models to change the different knobs of the cooling system, thus saving a significant amount of energy \cite{gao2014machine}. These applied use cases firmly attest to the feasibility of learning-based solutions in different aspects of resource management in Clouds.

\par Therefore, to deal with complexity of data centre infrastructures and dynamic nature of Cloud workloads, learning based resource management methods are crucial. Furthermore, to achieve significant energy efficiency, integrated resource management of computing and cooling system is necessary and manage the trade-offs between these two subsystems. In the next subsection, we describe the need for integrated resource management solutions in Cloud data centres. 

\subsection{Need for Integrated Energy and Thermal-aware Resource Management}
\begin{figure}[!t]
\centering 
\captionsetup{justification=centering}
\includegraphics[width=0.8\textwidth]{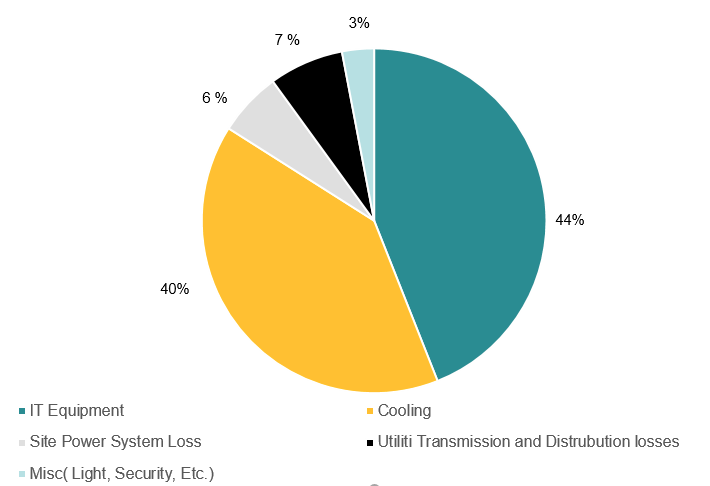}
\caption{Energy Distribution in Data Centres \cite{johnson2009data} }
\label{chap2-fig:energydist}
\end{figure}
Data centres hosts numerous subsystems, including IT/compute (compute servers, network, and storage equipment), cooling system, power distribution, and other facility-related subsystems. However, the majority of power is spent on computing and cooling systems. As shown in Figure\ref{chap2-fig:energydist}, computing and cooling system together account for  85\% of total energy consumption in a data centre, with each of them equally contributing \cite{johnson2009data}. 

\par Traditionally, cooling system management is left to the facility management team, and the computing system is managed by IT administrator individually. However, optimising a single system has an adverse effect on other systems. For instance, increasing resource utilisation in computing may create hotspots and thus increasing cooling energy cost. Hence, managing these subsystems separately leaves energy inefficiencies in data centre even though individually they are optimised for energy efficiency. The advancement in IoT and smart systems \cite{5958006} has enabled many mechanical systems associated with cooling to be managed or configured through software systems \cite{7805265} \cite{liu2016green} \cite{8073958}. Hence, it is imperative to apply resource management techniques holistically to optimise computing and cooling systems and avoid conflicting trade-offs between these two subsystems.

\subsection{Need for learning-based Resource Management Solutions}
The existing Resource Management Systems (RMS), from operating systems to large scale data centres, are predominantly designed and built using preset threshold-based rules or heuristics. These solutions are static and often employ reactive solutions \cite{bianchini2020toward}; they work well in the general case but cannot adjust to the dynamic contexts \cite{JeffMLforSystem}. Moreover, once deployed, they considerably fail to adapt and improve themselves in the runtime. In complex dynamic environments (such as Cloud and Edge), they are incapable of capturing the infrastructure and workload complexities and hence fall through. Consequently, the learning-based approaches built on actual data and measurements collected from respective DCS environments are more promising, perform better, and adapt to dynamic contexts. Unlike heuristics, these are data-driven models built based on historical data. Accordingly, these methods can employ proactive measures by foreseeing the potential outcome based on current conditions. For instance, a static heuristic solution for scaling the resource uses workload and system load parameters to trigger the scaling mechanism. However, this reactive scaling diminishes the users' experience for a certain period (due to the time required for system bootup and application trigger).
Consequently, a learning-based RMS enabled by data-driven Machine Learning (ML) model can predict the future workload demand and scale up or scale down the resources beforehand as needed. Such techniques are highly valuable for both users to obtain better QoS and service providers to offer reliable services and retain their business competency in the market. Moreover, methods like Reinforcement Learning (RL) \cite{JeffMLforSystem}. \cite{sutton1998introductionRL} \cite{sutton2018reinforcement} can improve RMS's decisions and policies by using monitoring and feedback data in runtime, responding to the current demand, workload, and underlying system status.

Machine learning-based RMS is more feasible now than ever for multiple reasons: (1)
AI techniques have matured and have proven efficient in many critical domains such as computer vision, natural language processing, healthcare applications, and autonomous vehicles; (2) Most distributed system platforms generate enormous amounts of data currently pushed as logs for debugging purposes or failure-cause explorations. For example, Cyber-Physical-Systems (CPS) in data centres already have hundreds of onboard CPU and external sensors monitoring workload, energy, temperature, and weather parameters. Such data is useful to build ML models cost-effectively; (3) the increasing scale in computing infrastructure and its complexities require automated resource management systems that can deliver the decisions based on the data and key insights from experience, to which AI models are ideal. In this paper, we study different techniques employed by researchers and practitioners for energy and thermal efficiency in cloud data centres.

\subsection{Contributions}
The  key contributions of this paper are summarised as below:
\begin{itemize}
    \item Describes need for integrated  and learning based solutions for  resource management  in Cloud data centres.
    \item Studies challenges associated with learning-based resource management methods.
    \item Proposes a taxonomy of different resource management techniques for energy and thermal efficiency based on an depth literature study
    \item Identifies future research directions for sustainable growth of Cloud Services.  
\end{itemize}

\subsection{Comparison to Existing Survey Papers}
Energy and Thermal efficient resource management of cloud data centres is a widely studied area by researchers. There have been many survey papers in this domain 
\cite{mittal-gpu-survey} \cite{orgerie-energy-ds} \cite{p2penergy} \cite{chana-energy-cloud} \cite{software-energy} \cite{power-energy-predictive-hpc} \cite{energy-cloud-fahime} \cite{green-energy=survey-dc} \cite{edge-cloud-energysurvey} \cite{tony-energy-cloud} 
\cite{hameed2016surveyresourceallocation} \cite{ahmad2015surveyconsolidation} \cite{thermal-awareshceduling} \cite{zhang2016towardsictandcooling}.  However, many of these survey  papers solely focus on energy efficiency of individual computing device types \cite{mittal-gpu-survey} and  study specific resource management aspects such as scheduling or resource allocation \cite{hameed2016surveyresourceallocation} \cite{thermal-awareshceduling}, and  consolidation \cite{ahmad2015surveyconsolidation}. While many other survey  papers focus on distributed systems such as peer-to-peer \cite{p2penergy} and also cloud computing systems \cite{chana-energy-cloud} \cite{tony-energy-cloud}. Some recent surveys are  focused on machine learning based solutions for various distributed computing systems \cite{power-energy-predictive-hpc}. Consequently, very few survey papers have been focused on  taxonomy and survey of energy and thermal efficiency \cite{gill2018taxonomy} \cite{zhang2016towardsictandcooling}. In contrast, this paper presents an holistic overview of data centres resource management for energy and thermal efficiency with detailed taxonomy and literature study. Moreover, we also highlight studies based on machine learning techniques which are finding real uses cases in modern resource management systems. Thus, our taxonomy and survey focusing on the integrated and learning based solutions brings new perspectives and key insights for resource management in Cloud data centres for energy and thermal efficiency.

\subsection{Article Organisation}
\par The rest of the paper is organised as follows: The brief background of Cloud computing,  its  energy consumption, and challenges associated with learning based resource management solution  is described in Section  \ref{sec:chap2-bg}. Section \ref{sec:chap2:taxonomyenergythermal} presents high-level taxonomy of energy and thermal aware resource management. Section \ref{sec:chap2:taxonomyenergy} describes existing methods based on taxonomy for energy management in data centre and Section \ref{sec:chap2:taxonomythermal} covers thermal management taxonomy and relevant solutions. The integrated resource management solutions for energy and thermal efficiency   are explained in Section \ref{sec:chap2:taxonomyjoint}. Then, Section \ref{sec:chap2coolingmanagement} describes different cooling managements systems in a data centre, including air and liquid cooling systems. 
Section \ref{Sec_Future} outlines the future research directions. Finally, Section \ref{chap2-Sec:summary} concludes the paper.

\section{Background}\label{sec:chap2-bg}

\subsection{ Cloud Data Centres and Energy Consumption}

\begin{figure}[!t]
\centering 
\captionsetup{justification=centering}
\includegraphics[width=\textwidth]{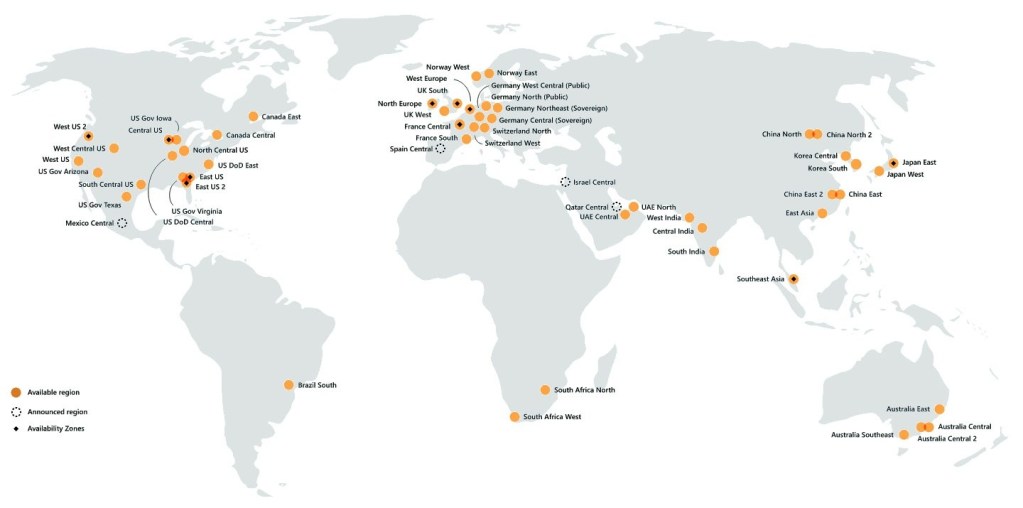}
\caption{Data Centre Locations of Microsoft  Azure  Cloud}
\label{chap1_Fig:dclocs}
\end{figure}

Cloud computing has seen tremendous growth in recent years. The transition from ownership-based on-premise IT infrastructure to subscription-based Cloud has changed the way computing services are delivered to end-users  \cite{buyya2009cloud} \cite{fox2009above}. Cloud computing's fundamental principle is to provide computing resources as utility services (e.g., water and electricity). It offers on-demand access to elastic resources with a pay as you go model based on  actual resource usage. This unique and flexible service delivery model ensures that individuals and businesses can easily access required computing services. 

\par Cloud computing services are broadly categorised into three types. First, the Infrastructure as a Service (IaaS) model offers computing, storage, and networking resources either in the virtual or physical form.  Second, the Platform as a Service (PaaS) model offers tools for rapid application development and deployment such as middleware platforms, Application Programming Interfaces (APIs), and Software Development Kits (SDKs).  Finally, Software as a Service (SaaS) model offers direct access to application software to the users, and the software is developed and managed by service providers completely. 

 Clouds have become application back-end and storage infrastructures for these modern IT services. Along with remote Clouds, recently, Cloud services are delivered from the edge of the network to satisfy Quality of Service (QoS) requirements for latency-sensitive applications such as autonomous vehicles, emergency healthcare services \cite{mahmud2018fog} \cite{satyanarayanan2017emergence}.  To seamlessly deliver services for applications and their users, Cloud computing uses massive network-based infrastructures. In particular, Data Centres (DCs) are the core and backbone infrastructure of this network system.  The DCs hosts thousands of servers, networking equipment, cooling systems, and facility-related subsystems to deliver reliable and uninterrupted services. By default,  Cloud workloads require a continuous, always-on, and 24$\times$7 access to its deployed services.   For instance, the Google search engine is expected to achieve an almost 100\% availability rate \cite{brin1998anatomy}. Similarly,  Amazon AWS witnesses thousands of Elastic Compute (EC2) instances created \cite{aws} in a day through their automated APIs, thus requiring massive geo-distributed DC infrastructures to support such critical demand. According to Gartner, by 2022, 60\% of organisations will use an external Cloud service provider \cite{gartner1}, and by 2024, Cloud computing alone accounts for 14.2\% of total global IT spending \citep{gartner2}.  

\par To cater for the demand of Cloud services, major Cloud service providers such as Amazon AWS\footnote{https://aws.amazon.com/}, Microsoft Azure\footnote{https://azure.microsoft.com/}, and Google Cloud\footnote{https://cloud.google.com/} are deploying a large number of hyper-scale data centres in multiple regions worldwide. A  snippet of Azure global data centre locations can be found in Figure \ref{chap1_Fig:dclocs} \cite{dcmap2}. Data centres have seen huge growth both in number and size. There are over 8 million data centres globally, from private small-scale to hypers-scale  DCs, and they are estimated to grow rapidly at 12\%  annually \cite{dcnumbers}. As their numbers and size grow, they are consuming an increasing amount of energy, resulting in massive energy challenges. DCs are power-hungry and require a continuous energy supply to power their computing, networking, and cooling systems. It is estimated that the DCs consume 2\% of global electricity generated \cite{USreport2016}. Furthermore, this massive energy consumption leads DCs to rely on fossil-fuel based or brown energy sources that hugely contribute to greenhouse gas emissions. DCs are responsible for emitting 43 million tons of CO\textsubscript{2}  per year and continues to grow at an annual rate of 11\% \cite{koomey2011growthCarbon} leaving high carbon footprints. Therefore, improving the Cloud data centre's energy efficiency is quintessential for sustainable and cost-effective Cloud computing.  

\begin{figure}[!t]
\centering 
\captionsetup{justification=centering}
\includegraphics[width=0.6\textwidth]{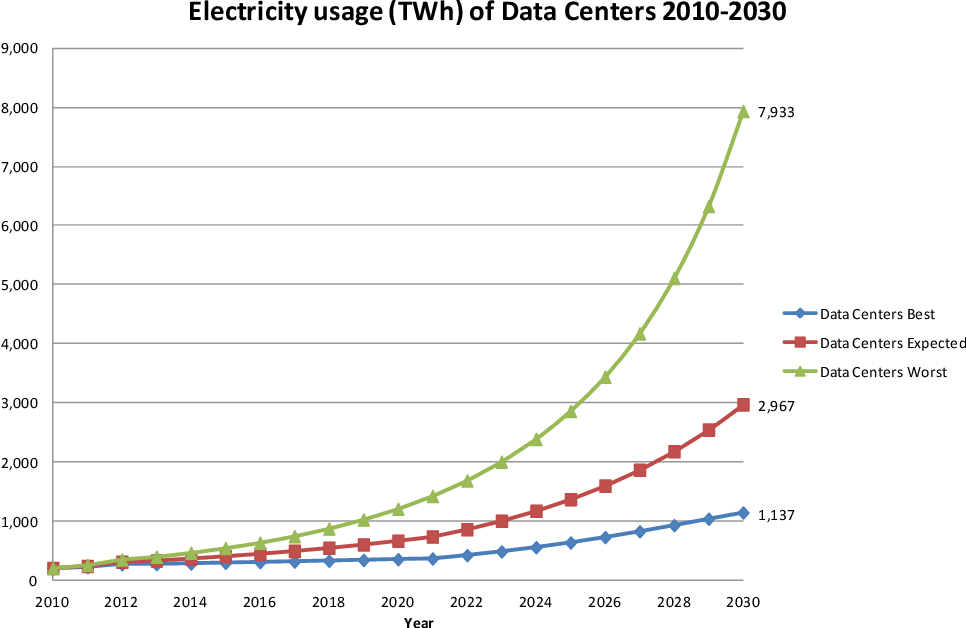}
\caption{Estimation of Data Centre Energy Consumption by 2030 \cite{andrae2015global}}
\label{chap1_Fig:dc2030}
\end{figure}

Cloud Data centres tremendous growth has introduced massive energy challenges. If necessary steps are not taken, data centres may consume up to 8000 terawatts of power in the worst case by 2030. However, if best practices are adopted across the Cloud computing stack, this massive energy consumption can be brought down to around 1200 terawatts \cite{andrae2015global} (see Figure \ref{chap1_Fig:dc2030}). To achieve this best-case scenario,  adopting energy-efficient practices into the various level of data centre resource management platforms (such as optimised use of computing and cooling resources) is necessary. Hence, it is of utmost importance to address this energy problem and achieve sustainability, both environmentally and economically.

\subsection{Challenges of learning-based Resource Management Solutions}
\subsubsection{Availability of Data}
The quality of data used to train the models determines the success of machine learning techniques. Also, this data should be available in large quantities with enough features covering all the aspects of environments \cite{cummins2017end}\cite{cano2019optimizing}. Within Cloud data centres, multiple challenges exist concerning the availability of such data. First, currently, different resource abstraction platforms collect the data at different granularity. The physical machine-level data from onboard sensors and counters is gathered and accessed by tools like Intelligent Platform Management Interface (IPMI), while at a higher abstraction level, middleware platforms collect data related to workload level, user information, and surrounding environmental conditions (temperature, cooling energy in the data centre). Also, network elements such as SDN controllers collect data related to network load, traffic, and routing. Unifying these data together and preprocessing it in a meaningful way is a complex and tedious task. The respective tools gather the data in a different format without common standards between them. Hence, building data-pipelines combining various subsystems data is crucial for the flexible adoption of ML solutions. Secondly, current monitoring systems collect data and push them into logging repositories to be used later for debugging. However, converting this data for ML-ready requires monotonous data-engineering. Hence, future systems should be explicitly designed to gather the information that can be directly fed to the ML models with minimal data engineering and preprocessing effort. Lastly, although several publicly available datasets provide workload traces, there are hardly any public datasets available representing various infrastructure, including physical resource configurations, energy footprints, and several other essential parameters (due to privacy and NDAs). Therefore, getting access to such data is a challenge and needs collaborative efforts and data management standards from the relevant stakeholders. Moreover, requiring standardised data formats and domain-specific frameworks \cite{portugal2016survey}.

\subsubsection{Managing the Deployment of Models}\label{sec:managedeploy}
Training ML models and inference in runtime needs an expensive amount of computational resources. However, one significant challenge is to manage the life cycle of ML models, including deciding how much to train, where to deploy the training modules in multi-tier computing architectures like Edge/Fog.  ML models tend to learn at the expense of massive computational resources consuming an enormous amount of energy. Therefore, innovative solutions are needed to decide how much learning is sufficient based on specific constraints (resource budget, time budget, etc.) and estimate context-aware adaptive accuracy thresholds of ML models \cite{toma2019adaptive}. To overcome this, techniques like transfer learning, distributed learning can be applied to reduce computational demands \cite{cano2019optimizing}. In addition, dedicated CPUs, GPUs, and domain-specific accelerators like Google TPU, Intel Habana, and FPGAs (Azure)  can carry out the inference.

\subsubsection{Non-Deterministic Outputs }
Unlike statistical models, which are analogous for their deterministic outputs, ML models are intrinsically exploratory and depend on stochasticity for many of their operations, thus producing non-deterministic results. For example, the cognitive neural nets, which are basic building blocks for many regressions, classification, and Deep Learning (DL) algorithms primarily rely on the principles of stochasticity for different operations (stochastic gradient descent, exploration phase in RL). When run multiple times with the same inputs, they tend to approximate the results and produce different outputs \cite{russell2002artificial}. This may pose a severe challenge in the Cloud systems,  where strict Service Level Agreements (SLAs) govern the delivery of services requiring deterministic results. For example, if a service provider fixes a price based on certain conditions using ML models, consumers expect the price to be similar all the time under similar settings. However, ML models may have a deviation in pricing due to stochasticity creating transparency issues between users and service providers. Many recent works have focused on this issue, and introduced techniques such as induced constraints in neural nets to produce the deterministic outputs \cite{lee2019gradient}. Yet, stochasticity in the ML model is inherent and requires careful monitoring and control over its output.

\subsubsection{Black Box Decision Making }
The ML models' decision-making process follows a completely black-box approach and fails to provide satisfactory justification for its decisions. The inherent probabilistic architectures and enormous complexities within ML models make it hard to evade the black-box decisions. It becomes more crucial in an environment such as DCS, where users expect useful feedback and explanation for any action taken by the service provider. This is instrumental in building trust between service providers and consumers. For instance, in case of a high overload condition, it is usual that the service provider shall preempt few resources from certain users at the expense of certain SLA violations. However, choosing which users' resources should be preempted is crucial in business-driven environments. This requires simultaneously providing fair decisions and valid reasons. Many works have undertaken to build the explanatory ML models (Explainable AI- XAI) to address this issue \cite{arrieta2020explainable}, \citep{gunning2017explainable}. However, solving this continues to remain a challenging task.

\subsubsection{Lightweight and Meaningful Semantics}
The DCS environment having heterogeneous resources across the multi-tiers accommodates different application services.  RMS should interact between different resources, entities, and application services to efficiently manage the resources. However, these requisites semantic models that represent all these various entities meaningfully. Existing semantic models are either heavy or inadequate for such complex environments. Therefore, lightweight semantic models are needed to represent the resource, entities, applications, and services without introducing the overhead \cite{bermudez2016iot}.

\subsubsection{Complex Network Architectures, Overlays, Upcoming Features}
Network architectures across distributed Clouds and telecom networks are evolving rapidly using software-defined infrastructure, hierarchical overlay networks, Network Function Virtualization (NFV), and Virtual Network Functions (VNF). Commercial Clouds like Amazon, Google, and Microsoft have recently partnered with telecom operators worldwide to deploy ultra-low latency infrastructure (AWS Wavelength and Azure Edge Zone, for example) for emerging 5G networks. The explosion of data from these 5G deployments and resource provisioning for high bandwidth, throughput, and low latency response through dynamic network slicing requires a complex orchestration of network functions \cite{zhang2017networkslice}.

In future Cloud systems, RMS needs to consider these complex network architectures, the overlap between telecom and public/private Clouds, and service function orchestration to meet end-to-end bandwidth, throughput, and latency requirements. These architectures and implementations, in turn, generate enormous amounts of data at different levels of the hierarchical network architecture. As different types of data are generated in different abstraction levels, standardised well-agreed upon data formats and models for each aspect needs to be developed.

\subsubsection{Performance, Efficiency, and Domain Expertise}
Many ML algorithms and RL algorithms face performance issues like a cold-start problem. Specifically, RL algorithms spend a vast amount of the initial phase in exploration before reaching their optimal policies creating an inefficient period where the decisions are suboptimal, even wholly random or incorrect leading to massive SLA violations \cite{cano2019optimizing}. RL-based approaches also face several challenges in the real world including (1) the need for learning on the real system from limited samples, (2) safety constraints that should never or at least rarely be violated, (3) the need for reward functions that are unspecified, multi-objective, or risk-sensitive, (4) inference that must happen in real-time at the control frequency of the system \cite{dulac2019challenges}. In addition, AI models are compute-heavy and designed with a primary focus on accuracy-optimisation resulting in a massive amount of energy consumption \cite{schwartz2019allen}. Consequently, new approaches are needed to balance the trade-offs between accuracy, energy, and performance overhead. Furthermore, current ML algorithms, including neural network architectures/libraries are primarily designed to solve computer vision problems. Adapting them to RMS tasks needs some degree of transformation of the way input and outputs are interpreted. Current AI-centric RMS algorithms transform their problem space and further use simple heuristics to interpret the result back and apply it to the RMS problems. Such complexities demand expertise from many related domains. Thus, newer approaches, algorithms, standardised frameworks, and domain-specific AI frameworks are required to adopt AI in RMS efficiently.

Despite the challenges associated, machine learning-based solutions provide many opportunities to incorporate these techniques into RMS and benefit from them. This paper explores different avenues where such techniques can be applied to manage Cloud data centres for energy and thermal efficiency.

\begin{figure}[!t]
\centering 
\captionsetup{justification=centering}
\includegraphics[width=0.8\textwidth]{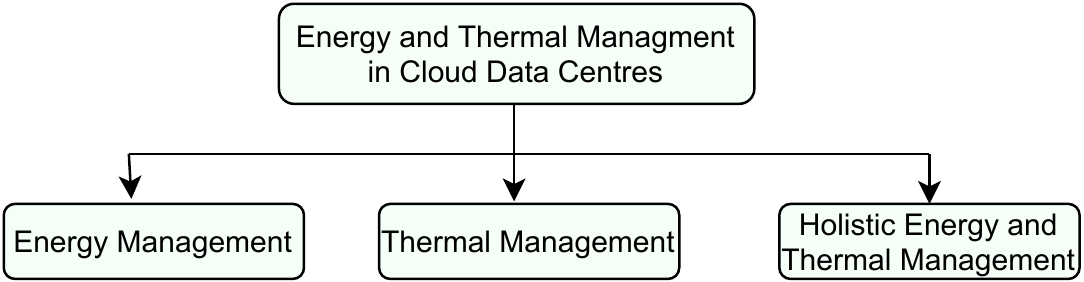}
\caption{A High Level Taxonomy of Energy and Thermal-Aware Resource Management Approaches}
\label{chap2_fig:highleveltaxonomy}
\end{figure}

\section{Taxonomy of Energy Thermal-aware Resource  Management in Cloud Data centres} \label{sec:chap2:taxonomyenergythermal}

As discussed earlier, cooling and computing are two main subsystems that contribute a significant amount of energy in data centres. In that regard, many works have focused on optimising these two systems with different optimisation methods.  Some works have also focused on the holistic optimisation of two systems by co-ordination between two subsystems, using techniques such as  power budget shifting and workload scheduling. A high-level taxonomy of these can be found in Figure \ref{chap2_fig:highleveltaxonomy}. The proposed solutions for energy and thermal efficiency can be    categorised into three types including (1) energy management; (2) Thermal management; and (3) holistic or integrated  energy and thermal management  of Cloud data centres. In this paper, we review existing research works in these three resource management categories. We  propose a taxonomy covering different optimisation techniques in each of these resource management categories. We focus on server level and data centre level  resource management solutions.

\section{Energy Management}\label{sec:chap2:taxonomyenergy}

\begin{figure}[!t]
\centering 
\captionsetup{justification=centering}
\includegraphics[width=0.8\textwidth]{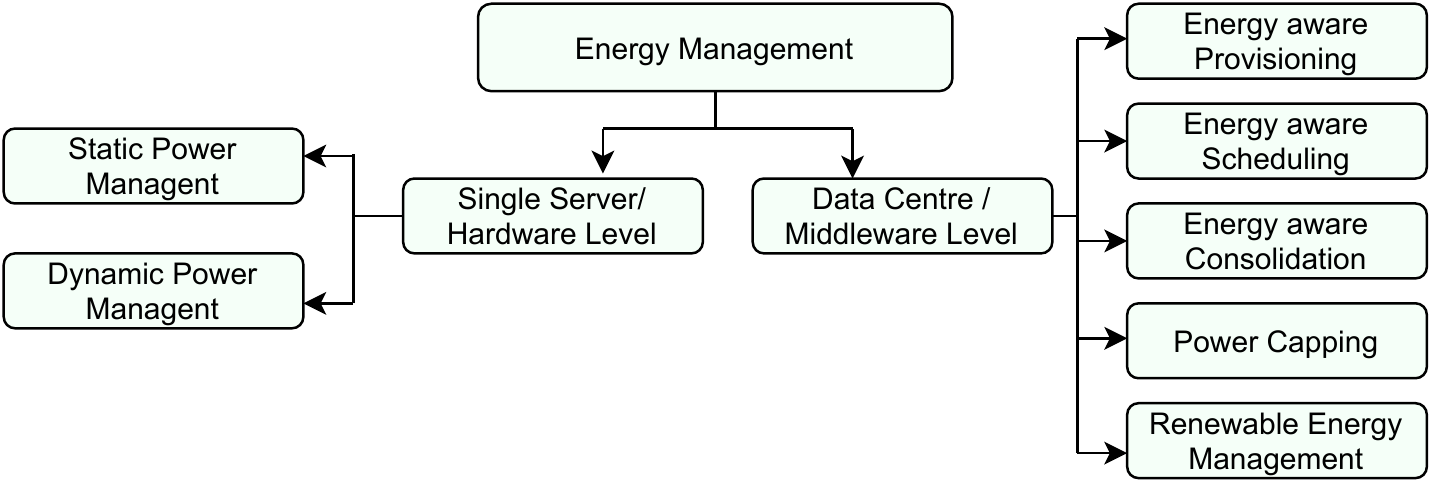}
\caption{ Taxonomy of Energy  Management in Cloud Data Centres}
\label{chap2_fig:energymanagementtaxonomy}
\end{figure}

Many researchers have focused on increasing the energy efficiency of data centres with various resource management techniques. These techniques cover from an individual server to geo-distributed data centres. Taxonomy on the data centre's energy management solutions is presented in Figure \ref{chap2_fig:energymanagementtaxonomy}. We categorise these solutions into two broad categories, i.e., single server level and data centre level solutions. Accordingly, we identify the essential techniques used in these two categories and briefly review their methods. 

\subsection{Server Level}

In a computing server, the CPU predominantly consumes a significant amount of energy. Modern rack-mounted data centre servers consume more than 1000 watts of power. Hence, managing this high power consumption is a challenging task. This server level power management has been mostly left to the operating system and its device drivers that communicate with underlying hardware signals and manage the server power.  Server level power management can be broadly categorised into two levels, static and dynamic power management. Static power management deals with minimising leakage power while dynamic power management deals with regulating active runtime power based on utilisation level.

\subsubsection{Static Power Management}
The silicon chip has static power consumption which is independent of the usage level. The static power mainly accounts for leakage of current inside active circuits. To some extent, static power consumption is unavoidable; however, it can be minimised with better design and processes. There are many solutions from a lower level from circuit level, and architectural techniques \cite{venkatachalam2005power}. The general approach in managing leakage is with different sleep states of CPUs when the system is idle. For instance, Intel X86 architecture has (C0-C4) sleep states indicating C0 is an active state while C4 is a deep sleep state where most of the CPUs' components are turned off to avoid static power consumption.

\subsubsection{Dynamic Power Management}
A large part of silicon chip-based computing elements either in CPU or GPUs spend on dynamic power. Dynamic power represents runtime energy based on workload utilisation level. CPUs operate at a different frequency to regulate the dynamic power. If the operating frequency of a CPU is highest, then its dynamic power consumption will also be higher. The frequency is regulated based on utilisation level and workload requirements to increase their speedup.  Dynamic Voltage Frequency Scaling (DVFS) is a popular technique to regulate the dynamic power in modern systems  \cite{Bridges2016}. The dynamic power can be defined as below:
 \begin{equation}
  \label{chapt2eq:dynpowerpower}
            P_{dynamic} \hspace{.2cm} \alpha  \hspace{.2 cm} V^{2}F
\end{equation}

\par In Equation \ref{chapt2eq:dynpowerpower}, $F$ is the frequency, and  $V$ is the supply voltage to the processor.
 Based on the frequency, the voltage is regulated, and some frequency ranges usually have a similar.  If a CPU should be at its highest speed or frequency should be set to a higher level, thus consuming more power. The operating system scales frequency based on its workload and application demands in runtime.  
 
\par There are many solutions proposed that intend to optimise energy efficiency through  DVFS techniques at the data centre level. These solutions include DVFS-aware VM scheduling and consolidation \cite{von2009power} \cite{arroba2017dynamic}, placement of application based on DVFS capabilities \cite{safari2018energy}, data centre level task scheduling by synchronising the frequency scaling among multiple machines \cite{wang2017dvfs}. 
ML-based techniques have also been explored recently in DVFS optimisations. Authors in \cite{park2017ml} proposed ML-based CPU and GPU DVFS regulator for compute-heavy mobile gaming application that coordinates and scales frequencies with performance and energy improvements. 

\subsection{Data Centre Level}

A significant amount of energy efficiency can be achieved when data centre level platforms incorporate energy-efficient resource management policies. Distributed data centre applications span hundreds of machine in geo-distributed data centres, hence, providing energy efficiency holistically across data centre resources and applications is more feasible and yields better results. In this section, we discuss important techniques for data centre level energy-efficient solutions. 

\subsubsection{Energy-aware Provisioning}
Cloud data centre offer computing resources in terms of Virtual Machines (VMs) or containers. Allocating the required amount of resources for the application need is vital to satisfy the SLAs. However, overprovisioning of resources may yield higher energy consumption, and monetary cost to the users while underprovisioning will potentially violate the SLAs. Many researchers have proposed energy-aware resource provisioning techniques.
Authors in \cite{7276993} investigated energy-aware resource allocation for scientific applications. The proposed system  EnReal leverages the dynamic deployment of VMs for energy efficiency. Similarly, Li et al.  \cite{7918312} proposed an iterative algorithm for energy-efficient VM provisioning for application tasks.  Beloglazov et al. \cite{beloglazov2012energy} propose various heuristic algorithms for resource allocation policies for VMs defining architectural principles.

\par Some researchers have also proposed data-driven methods for resource provisioning. Mehiar et al. \cite{7111351} offered clustering and prediction based techniques; they used K-means for workload clustering and stochastic Wiener filter to estimate the workload level of each category accordingly allocate resources for energy efficiency. Recently Microsoft has proposed Resource Control (RC) \cite{bianchini2020toward}, where they trained ML models to output predictions like VM lifetime, CPU utilisation, maximum deployment of VMs. These predictions use various resource management problems for better decision-making, including resource provisioning with the right container size for applications. 

\subsubsection{Energy-aware Scheduling}
Scheduling is a fundamental and essential task of a resource management system in Cloud data centres. It addresses the following question, given an application or set of  VMs (considering application runs inside these isolated VMs), when and where to place these  VMs/application among available physical machines. This decision depends on several factors, including application start time, finish time, and required SLAs. In addition,  workload models, whether an application is a long-running (24 $\times$7) web application, or a scientific workflow model of which it's tasks need to be aware of precedence constraints, or applications based on IoT paradigm that is predominantly event-driven. Although one can optimise numerous scheduling parameters, many recent studies have focused on energy optimisation as a priority in Cloud data centre scheduling.

\par  Chen et al. \cite{chen2015towards} propose energy-efficient scheduling in uncertain Cloud environments. They propose an interval number theory to define uncertainty, and a scheduling architecture manages this uncertainty in task scheduling. The proposed PRS1 scheduling algorithm based on proactive and reactive scheduling methods optimises energy in independent tasks scheduling. Similarly, Huang et al. \cite{6217511} investigate energy-efficient scheduling for parallel workflow application in Cloud. Their  EES algorithm tries to slack non-critical jobs to achieve power saving by exploiting the scheduling process's slack room. Energy-efficient scheduling using various heuristics for different application model has been widely studied topic in literature \cite{ding2015energy} \cite{7037687} \cite{ghribi2013energy}. 

\par Machine learning-based solutions are also explored in data centre scheduling focusing on energy efficiency. Some solutions rely on predictive models and then use them in scheduling algorithms, while other techniques model scheduling as a complete learning-based problem using t Reinforcement learning (RL). Berral et al. \cite{berral2010towards}, adopt many ML-based regression techniques to predict CPU load, power, SLAs and then use these in scheduling for better decisions. These solutions still use some level of heuristics with integrated prediction models. However,  RL-based scheduling is designed to learn and take actions in a data centre environment without external heuristics. Cheng et al. \cite{cheng2018drl} proposed DRL-based provisioning and scheduling for application tasks in the data centre. 

\subsubsection{Energy-aware Consolidation}
Cloud data centre are designed to handle the peak load to avoid potential SLA violation or overload conditions. Hence, the resources are oversubscribed to manage such an adverse situation. However, this oversubscription leads to resource underutilisation in general. It is estimated that Cloud data centres utilisation level is around 50\% on average. Under utilisation of resources is the main factor in the data centre's energy inefficiency as idle or lower utilised servers consume significant energy (up to 70\% \cite{beloglazov2011taxonomy}). Thus, it is necessary to manage workloads under such oversubscribed and underutilised environments. To that end, consolidation has been a widely used technique to increase energy efficiency. It aims to bring the workloads (VMs and containers) from underutilised servers and consolidate them on fewer servers, thus allowing remaining servers to be kept in sleep/shut down mode to save energy. Many challenges exist in consolidation, including maintaining VM-affinity, avoiding overutilisation, minimising SLA violation, and reducing application downtime due to workload migrations. 

Beloglazov et al. \cite{beloglazov2012energy} proposed various heuristics to consolidate the workload and answer the question, including which VMs to migrate, where to migrate and when to migrate to reduce potential SLA Violation.  Many other solutions have broadly focused on energy efficiency along with optimising different parameters (cost reduction, failure management, etc) while consolidating workloads in the data centre \cite{piraghaj2015framework} \cite{ferdaus2014virtual} \cite{sharma2019failure}.

\par Data-driven solutions are  predominantly used in consolidation \cite{hieu2017virtual} \cite {hsieh2020utilization}. Hsieh et al. \cite {hsieh2020utilization} studied  VM consolidation to reduce power cost and increase QoS. They predict the utilisation of resources using the Gray-Markov-based model and use the information for consolidation. Similarly, the authors \cite{hieu2017virtual} also use prediction for consolidation. They predict memory and network usage and perform consolidation of VMs in a data centre along with CPU. Few researchers have also used RL in energy-aware consolidation \cite{farahnakian2014energy}  \cite{basu2019learn}. Basu et al. \cite{basu2019learn}proposed Megh--- a system that learns to migrate VMs in the data centre using RL. It proposes the dimensionality reduction technique using dimensional polynomial space with a sparse basis to minimise the state-space in their problem. Their system has shown that it achieves better energy efficiency and cost reduction compared to existing heuristics. 
\subsubsection{Power Capping}
Data centres are designed to handle the peak power consumption based on the workload and cooling system requirements. Hence, in general, data centres are under-provisioned with power. This power capping on data centre servers restricts the amount of energy available to individual servers even though they can consume their maximum limit, thus providing the required speed for workloads \cite{bhattacharya2013need}.  Managing resources and workload effectively in these power-constrained environments is necessary. It is essential to avoid power inefficiencies in limited power allocated across servers to achieve power proportional computing \cite{petoumenos2015power}. 

\par In this regard, different power capping mechanisms at the Cloud data centre level are studied. The authors \cite{azimi2017fast} proposed a fast decentralised power capping (DPC) technique to reduce latency and to manage power at the individual server. Dynamo \cite {wu2016dynamo} is the power management system used 
by Facebook data centres, which has hierarchical power distribution. The lowest level leaf controller regulates power in a group of servers. This leaf controller based on a high-bucket-first heuristic determines the amount of energy to be reduced in each server to meet the power cap limits to which it is constrained. It also considers workload priorities and avoids potential performance degradation due to its power capping.  Some researchers have investigated controlling peak power consumption \cite{lefurgy2008power}by designing feedback controller, which periodically reads system-level power and configures highest power state of servers keeping server within its power budget. Authors in  \cite{gandhi2009optimal} studied optimal power allocation in servers, which accounts for several factors including power-to-frequency, the arrival rate of jobs,  maximum and minimum server frequency configuration. They have shown that allocating full power may not always result in the highest speed as expected. Some techniques have also explored enabling data centre service providers to dynamically manage the power caps by participating in an open electricity market and achieve cost and energy efficiency \cite{6691107}. However, due to close interconneciton between power capping effect on CPU speed, thermal dissipiation and also presence of heterogenety in servers and workloads, data centre level power capping  workload management is a difficult task to achieve \cite{zhang2016maximizing} as compared to other energy efficiency methods that are  discussed in this paper.  

\subsubsection{Renewable Energy Management}
Data centres consume colossal energy and contribute significantly to greenhouse gas emissions ($CO_2$). Data centre service providers continuously increase renewable or green energy (solar, wind) usage with minimal carbon footprints to decarbonise the data centres. However, green energy usage in the data centre is extremely challenging due to its intermittent nature availability. In contrast, the Cloud data centre needs an uninterrupted power supply since Cloud workloads tend to run $24\times7 $. Therefore, managing workloads under the uncertain availability of renewable energy is a challenging research problem.

Several resource management techniques explored maximising renewable energy in data centres.  They include workload shifting and placement across geo-distributed data centres \cite{xu2020managing} \cite{7937820} \cite{6678925} based on their carbon efficiency. Besides, delaying job execution if an application can tolerate the QoS \cite{goiri2012greenhadoop} and job dispatching or load balancing workloads to match the available renewable energy at different data centres \cite{zhang2011greenware} are some popular techniques in this regard. 

Machine learning-based algorithms are promising in renewable energy management, as predicting the available green energy based on an environmental condition is crucial in workload management \cite{lai2020survey}. Along with prediction models, RL methods are also used to solve optimisation problems in increasing green energy usages in data centres \cite{gao2020smartly}. 

\subsection{Summary of Energy Management in Data Centres}

To achieve significant energy efficiency in data centres, we need algorithms and software systems that manage resources and workloads across different computing layer. In addition to the energy management techniques we discussed in this section, researchers propose various other solutions. The proposed solutions cover designing more energy-efficient processor architecture, building middleware platforms that manage resources efficiently, and finally including energy efficiency in the software development process, itself \cite{capra2012software} \cite{pinto2017energy}. As data centre systems' complexities increase, machine-learning-based solutions are becoming predominant that either aid externally for different algorithms or directly taken action if modelled accordingly.

\section{Thermal Management}\label{sec:chap2:taxonomythermal}

Thermal efficient resource management in the data centre is vital to increase energy efficiency.  It is important to  manage peak temperature \cite{el2012temperature}. Every degree increase in data centre peak temperature costs millions of dollars in operational cost \cite{torell2015unexpected} as the cooling system's thermal load drastically increases.   Furthermore,  increased temperature affects the cooling cost and further decreases the system's reliability due to failures under high thermal conditions. It is essential to understand that the data centre's peak temperature and the cooling system setpoint temperature are different (also called supply air temperature). If the data centre's peak temperature increases, supply air temperature needs to be set to a lower value, requiring higher cooling energy. Hence, to solve these problems, a workload management system should be aware of such trade-offs, and resources should be managed holistically.  The data centre workloads should be managed to reduce energy consumption and keep the peak temperature of the data centre within the recommended thresholds, thus keeping cooling energy cost minimum.

\par
\begin{figure}[!t]
\centering 
\captionsetup{justification=centering}
\includegraphics[width=0.8\textwidth]{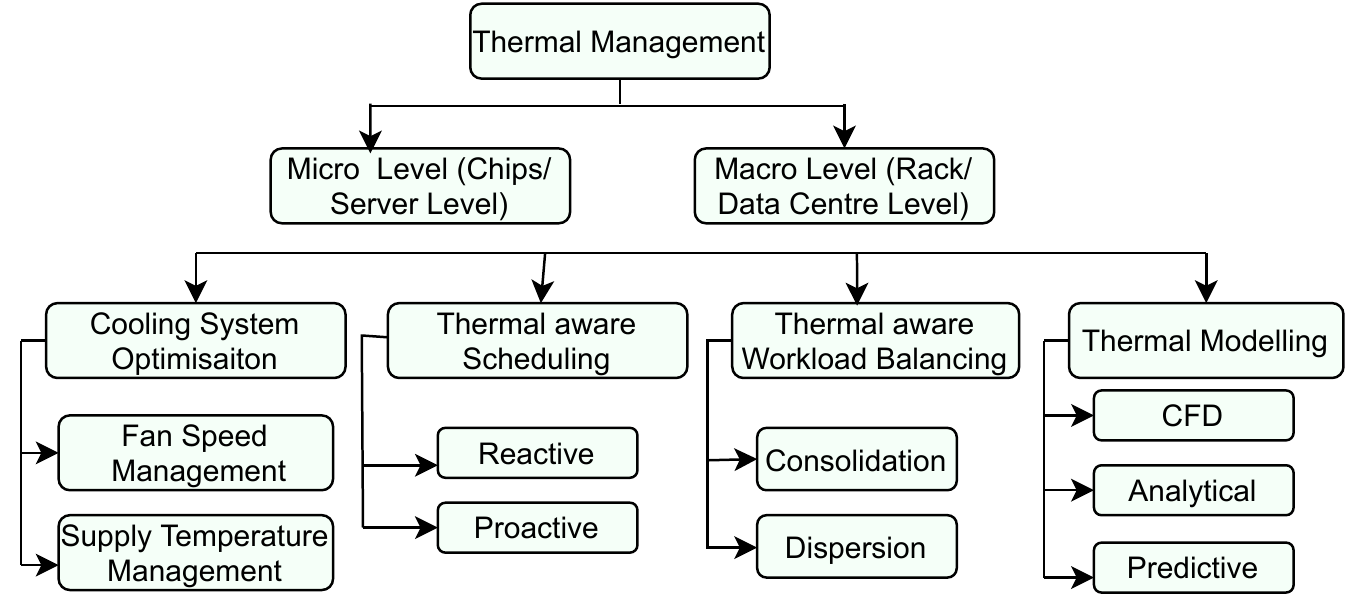}
\caption{ Taxonomy of Thermal  Management in Cloud Data Centres }
\label{chap2_fig:thermalmanagementtaxonomy}
\end{figure}

 Similar to energy management, thermal management techniques span from an individual server to data centres. A taxonomy on thermal management solutions is presented in Figure \ref{chap2_fig:thermalmanagementtaxonomy}. This section categorises these techniques into two broad categories, i.e., micro-level or single server level and macro-level or data centre level thermal management techniques. We describe and review essential approaches used in these two categories.

\subsection{Server Level}
Computing servers consume an enormous amount of energy and dissipate this energy as heat. It is crucial to keep processor or CPU temperature within the threshold limit to avoid damage to the processor's silicon components, thus permanently producing catastrophic device failures. Modern rack servers reach peak temperature up to 90-100 $\celsius$.  In reality, the processor speed of servers is limited by their thermal management capacity. Generally, onboard fans are responsible for taking out heat from the server cabinet to the outside ambient environments in data centres. 

\par Like DVFS in energy management, its corresponding thermal dissipation is regulated in servers by controlling the amount of power consumed. Dynamic Thermal Management (DTM) \cite{shin2010energy} is a popular thermal management technique at the individual server level which regulates Multiprocessors Systems-on-chip (MPSoCs) performance, power consumption, and reliability. This is controlled at the operating system level by closely communicating with underlying hardware interfaces. If a server's temperature is potentially exceeding the predefined threshold, the operating system takes major by employing thermal throttling mechanisms that reduce the energy consumption, thus reducing the CPU speed.  Moreover, techniques like application scheduling \cite{ayoub2011temperature} \cite{sheikh2012overview}, optimal onboard fan speed configuration \cite{wang2009optimal} techniques are employed for energy and thermal efficiency at server level. \par  Machine-learning based solutions are recently used to optimise temperature management at individual server level \cite{pagani2018machine}. For instance, Iranfar et al.    \cite{iranfar2020dynamic}  investigated how to proactively estimate the required number of active cores, operating frequency, and fan speed. Accordingly, the system is configured to achieve reduced power consumption. 

\par  The server-level thermal management involves solutions including processor architecture design, manufacturing technology and resource management solutions within the operating system, including server fan control and others. As our focus is entirely on data centre solutions, we do not delve into server-level thermal management.
\subsection{Data Centre Level}
 A typical large scale data centre hosts thousands of servers.  Data centre servers are arranged in rack-layout,  where each rack (e.g., standard 42U rack) can accommodate 10-40 rack blade servers based on vendor-specific dimensions. This high density of equipment makes the data centre one of the highest-energy density physical infrastructures. Disipiated heat from these rack server can result in the data centre ambient temperature reaching extremely high. Thus, cooling systems in data centres make sure that data centre temperature is within the threshold. Many approaches exist optimising different parameters to reduce cooling energy. In this section, we review and describe data centre level thermal management techniques. 

\subsubsection{Cooling System Optimisation}
Traditional rack layout data centres have a Computer Room Air Conditioning (CRAC)  cooling system that blows cold air to the racks across data centre (more details of cooling technologies can be found in Section \ref{sec:chap2coolingmanagement}). The entire cooling system efficiency requires multiple parameters to be configured in the design and operational phase. In the design phase, efficiency can be increased by better physical layout and vent designs to reduce heat recirculations. While runtime cooling energy efficiency can be increased by fine-tuning the fan speeds of CRAC systems and cold air supply temperature which mainly determines the cooling system energy consumption \cite{zhang2016towards} \cite{khalaj2017review} \cite{nadjahi2018review}. In this section, we focus on runtime cooling system optimisation. 
\\
\textbf{\textit{Fan Speed Management:}} Within the CRAC system,  fans are used to regulate the airflow rate within the data centre. It is important to note that these fan speeds are separate from the onboard server's fan equipped to eject heat from CPU to the outside of the server cabinet. Increasing airflow require higher fan speeds, thus consuming more energy. Hence, regulating fan speed optimally can save a significant amount of cooling power. However, this depends on the status of the data centre, and its temperature level.  Many researchers have proposed solutions to optimally configure the CRAC's fan speed based on cooling load  \cite{tian2019energy} \cite{zhang2016towards} by monitoring thermal load in the data centre and accordingly varying fan speeds dynamically to reduce energy consumption.
\\  
 \textbf{\textit{Supply Temperature Management:}} CRAC system blows cold air to racks through vented floor tiles in the data centre to take out dissipated heat.  Passing colder air requires higher energy consumption as chiller's in CRAC consumes energy to supply cold air. Hence, the inaccurate configuration of supply air temperature significantly affects cooling energy cost in the data centre.  For a safer operation, the American Society of Heating, Refrigerating and Air-Conditioning Engineers (ASHRAE) \cite{ASHRAE},  recommends supply air temperature in the data centre to be in the range of  17-27 $\celsius$.  Thus, it is beneficial to set the supply temperature closer to 27 $\celsius$. However, most data centres are overcooled as supply temperature in the data centre is set to much lower temperature conservatively, leaving energy inefficiencies in the cooling system. Setting a higher supply air temperature requires careful handling of peak temperature in data centres. 

Many solutions have been proposed to raise the supply air temperature. Zhou et al. \cite{zhou2011holistic} have shown that significant power saving can be achieved when the workload is managed efficiently and allowing supply air temperature to be increased. In essence, to raise supply air temperature, the data centre peak temperature should be minimised. It can be done through various means, including thermal aware workload scheduling, and avoiding thermal imbalance in the data centre. 
\subsubsection{Thermal-aware Scheduling}
Workload scheduling in the data centre has a significant effect on cooling system efficiency \cite{chaudhry2015thermal}. If workload scheduling strategy results in peak temperature in the data centre, it generates a higher thermal load, thus increasing cooling cost. To address this, many researchers have proposed thermal-aware scheduling methods in Cloud data centres. Some solutions are proactive, which intends to avoid adverse temperature effects beforehand. In contrast, some scheduling policies follow reactive approaches. If a temperature violation is found, workloads are rescheduled to other nodes; however, the reactive scheduling method may result in higher QoS violation for application due to rescheduling and migration.  Mhedheb et al. \cite{mhedheb2013load} investigated load and thermal aware scheduling in Cloud that optimises temperature and load while scheduling tasks in data centres. Sun et al. \cite{sun2017spatio} proposed thermal-aware scheduling of HPC jobs. They have used analytical models to estimate server temperature and model heat recirculation in the data centre. Proposed thermal aware job assignment heuristics have shown that increased performance with thermal balancing. Furthermore, authors in \cite{6690176} have further extended thermal aware batch job scheduling across geo-distributed data centres.

\par Many of the existing works have employed machine-learning-based techniques in thermal-aware scheduling. Wang et al.  \cite{wang2011task} proposed  Artificial Neural Networks (ANN)-based temperature prediction model and using it for task prediction in data centres. The results have shown that machine learning models are capturing the thermal phenomenon in a data centre. 
\subsubsection{Thermal-aware Workload Balancing}
In Cloud data centres, thermal agnostic placement of workload triggers adverse temperature effect. Hence, balancing the workloads thermal efficiently yields better efficiency, consolidation and workload dispersion are two popular techniques in workload balancing. 
\newline \textbf{\textit{Worklaod Consolidation:}} Consolidation is a widely used technique to optimise a computing system's energy consumption. However, aggressive consolidation leads to the creation of hotspots that further increases cooling cost. Hence, thermal-aware consolidation is necessary to balance the computing and cooling system energy consumption. Many researchers have proposed many solutions for this \cite{lee2012vmap} to balance the temperature response due to workload placement.
\newline \textbf{\textit{Workload Dispersion:}} Opposite to consolidation, the workload dispersion technique aims to spread out workloads evenly across the data centre's servers \cite{shamalizadeh2013optimized}. It has shown to be thermal efficient workload management as it minimises temperature in a data centre, avoiding servers to reach peak utilisation. Although it minimises peak temperature, it significantly increases the computing system energy due to resource underutilisation. Hence, there should be a balance between consolidation and workload dispersion techniques to achieve cooling system efficiency. 

\subsubsection{Thermal Modelling}
Thermal modelling in data centre plays a vital role in resource management. Thermal modelling includes capturing thermal behaviour in a data centre and accurately estimating server temperature. Thermal models that predict accurately and fastly are useful aids in scheduling, configuring cooling system and other resource management techniques. However, temperature prediction is a difficult problem. Server ambient temperature in a data centre depends on multiple factors including  CPU heat dissipation, inlet temperature and complex heat recirculation effects. There are mainly three types of thermal modelling techniques in data centres: (1) Computational Fluid Dynamic (CFD)-based models; (2) Analytical models; and (3) Predictive models.
\\
\textbf{\textit{CFD:}} The CFD models accurately captures the room layouts, heat recirculation effects and accurately estimates temperature in the data centre \cite{choi2008cfd} \cite{5224189} \cite{CFDsecond}. However, they are computationally expensive, and even a single calibration requires models to be run for multiple days. Hence, they are incapable of using resource management systems that require for their Fast online decisions. 
\\
\textbf{\textit{Analytical:}} These models depends on modelling data centre and workloads based on mathematical frameworks \cite{tang2008energy} \cite{sun2017spatio}. They represent cooling, computing and workload elements with formal mathematical models and build a framework to establish relationships between all elements \cite{sun2017spatio}. Although they are fast in temperature estimation, the accuracy is compromised due to their rigid static models. 
\\
\textbf{\textit{Predictive:}}
ML-based models use actual measurement data from the data centre to predict the accurate temperature of the server. These data-driven models, once trained, are accurate,  and quickly deliver the results in runtime. Moreover, they can automatically model physical layout, air conditioning and the heat generated by Cloud data centres. Unlike CFD's where each of these needs to be modelled explicitly, this is a huge benefit. To that end, Wang et al. \cite{wang2011task} proposed a server temperature prediction model using the Artificial Neural Network (ANN) based ML technique, results have shown that it can accurately predict the temperature in data centres. In addition, some studies have explored using machine learning models to identify temperature distribution \cite{tarutani2015temperature}, and to predict server inlet temperature \cite{lloyd2018data}.

\par The drawback of the data-driven model is that the model is only applicable to the data centre where the data is collected from.   This means data need to be collected for each data centre extensively.  However, this is not a massive disadvantage as such data need to be collected to monitor the data centres' health.

\subsection{Summary of  Thermal Management in Data Centre}
Efficient thermal management in a data centre is essential for achieving energy efficiency and guaranteeing system reliability. In this section, we reviewed various thermal management solutions spanning individual server to data centre level methods. Compared to energy management, machine-learning-based approaches in thermal management is limited or less explored. However, there exist vast opportunities to incorporate learning-based solutions across thermal management stack in Cloud data centres.

\section{Integrated Energy and Thermal Management}\label{sec:chap2:taxonomyjoint}
Traditionally cooling system and computing systems are optimised individually. However, these two subsystems in the data centre are closely interdependent and optimising one system often have a counter effect on others. Hence, the joint optimisation of two subsystems is beneficial. 
Integrated management of both computing and cooling energy is a challenging task that requires capturing complex dynamics of data centre workloads and physical environments.  
Many solutions have been proposed including workload scheduling and cooling system optimisation as a multiobjective optimisation problem and accordingly configure different parameters to minimise energy consumption \cite{7438750} holistically. Other techniques include CRAC fan speed management by interplaying with IT load and its heat dissipation, configuring supply air temperature, and distributing the workload to minimise peak temperature, among many others. 

\par Wan et al. \cite{7932106} studied holistic energy minimisation in data centres through a cross-layer optimisation framework for cooling and computing systems. This energy minimisation problem is formulated as a mixed-integer nonlinear programming problem. To solve this problem, the authors proposed a heuristic algorithm called  JOINT, that dynamically configures parameters  (such as server frequency, fan speed, and CRAC supply air temperature) based on workload demand and minimises computing and cooling system energy holistically. 

\par Li et al. proposed \cite{6258012}  joint optimisation of computing and cooling systems for energy minimisation in data centres by modelling IT systems interactions (load distributions) and its corresponding thermal behaviour, i.e., heat transfer. The proposed analytical models for load distribution across rack servers to minimise computing and cooling system energy, thereby configuring different knobs of two systems while ensuring required throughput and resource constraints of workloads.

\par Power budget shifting is another important resource management techniques in Join optimisation of these two systems. Using available power to trade between two systems in runtime can increase energy efficiency and resource utlisaiton.  PowerTrade \cite{ahmad2010joint} is a technique that trades-off data centre computing system's idle power and cooling power with each other to reduce total power. Over provisioning is necessary for such condition to accommodate extra workload and use excessive power obtained. 

\par Machine learning-based techniques have also been explored in joint optimisation of computing and cooling systems. Recent advancements in RL have made it possible to learn different policies by interacting with the environments and learning from experience. RL techniques can be more adaptive and automatically understand the policies. Ran et al. \cite{ran2019deepee} used DRL and designed hybrid action space that optimises the IT system and the airflow rate of the cooling system. Furthermore, the proposed control mechanism coordinates both the IT system's workload and cooling systems for energy efficiency.   Similar techniques can be found in other studies \cite{mirhoseini2017device} \cite{cheng2018drl}. Careful design of state management, action, and rewards are important for applying RL techniques to data centres' holistic energy management. 

\section{Cooling Management Technologies in Data Centre} \label{sec:chap2coolingmanagement}
When servers/IT equipment uses electricity for their operations, the electrical energy is transferred as heat. This heat will be drawn across the server cabinet by the rear-mounted server fans within allowing heat to transfer from the server's components to the outside ambient environment. Many technologies are employed to take out this heat from the data centre environment and keep the data centre's operational temperature within its threshold. These cooling technologies can be broadly categorised into two categories, including air and liquid cooling technologies. 
\subsection{Air Cooling}
Air cooling is a widely used data centre cooling technologies due to its inexpensive and flexible design and operational conveniences. In rack-layout based data centres,  the dissipated heat from servers is extracted from the cooling system's environment. The \textbf{Computer Room Air Conditioning} (CRAC)  is a cooling system responsible for monitoring and managing the temperature in the data centre \cite{fakhim2011cooling}. The  CRAC blows cold air through the perforated tiles under the racks of a data centre. The cold air passes from the bottom to the top of the rack taking out the dissipated heat from rack equipment sand this hot exhaust air is pushed to the intake of the CRAC units to the ceiling of the room where it is taken out of the room. This allows separating hot exhaust air from the cold inlet air. The CRAC unit then transfers the hot exhaust air via a coil, to a fluid using refrigerant.

Many data centre also equip  \textbf{Computer Room Air Handler (CRAH)}, where chilled water is used as fluid \cite{fakhim2011cooling}. These fluids remove the heat from the data centre environment.  The CRAC/CRAH continuously blow cold air using constant-speed fans, and this return cold-air temperature also called inlet temperature. It is configured to manage the dynamic thermal threshold in the data centre. It directly controls the cost of cooling in general. Lower the inlet temperature higher will be the cooling energy cost due to increased energy required to transfer the lower temperature air from CRAC/CRAH. The  American Society of Heating, Refrigerating and Air-Conditioning Engineers (ASHRAE) \cite{ASHRAE}, a leading technical Committee in cooling system technology recommends that the device inlet be between 18-27°C for the safe operation of the environment. The design goal of any data centre operators will be to provide the inlet temperature close to 27 °C to reduce the cooling cost. However, the safer operation threshold should be maintained while configuring this parameter. Many works have looked into optimising this parameter using different techniques by minimising the peak temperature \cite{tang2008energy}  by balancing the workloads \cite{ilager2020thermal} and optimally configuring other parameters \cite{lee2015proactive}  of the cooling system.

Some modern system also use \textbf{evaporative} \cite{porumb2016review} and \textbf{air side economisers/ free cooling} techniques \cite{zhang2014free}. In the evaporative technique,  instead of fluid refrigerant, the hot air carried from the data centre is directly exposed to water. Water evaporates, taking out the heat from the hot air. Cooling towers are employed to dissipates the excess heat to the outside atmosphere. However, it doesn't require expensive  CRAC or CRAH units but needs a large amount of water, a limiting factor in many data centre locations. On the other hand, air side economisers or free cooling methods use outside free air for direct cooling instead of depending on the fluids to cool down the hot air extracted from CRAC/CRAH. This saves a huge amount of cooling cost. Nonetheless, these techniques vastly depend on the weather and geographical condition where the data centre is located, and thus they are used in limited computing infrastructures in practice. 

\subsection{Liquid Cooling}
The recent advancement in the data centre cooling technology has seen the adoption of liquid cooling as it is more efficient than air cooling in general \cite{liquidcooling}. The liquid cooling system also effectively avoids heat mix up and heat re-circulation issue, which is a common problem in air cooling techniques.

 In \textbf{direct liquid cooling} system,  liquid pipes are used to deliver liquid coolant directly to the heat sink present in the server's motherboards. The dissipated heat from the server is extracted to heat the chiller plant from these pipes, where the chilled water loop takes out the heat extracted from servers.  

 \textbf{Immersion cooling}. The computing system (servers and networking equipment are directly immersed in a non-conductive liquid. The liquid absorbs the heat and transfers it away from the components \cite{capozzoli2015cooling}. In some cases, equipment is arranged in isolated cabinets and immersed in tanks or cabinets are directly immersed in natural water habitats such as lakes/ocean. For instance, Microsoft has tested underwater data centre with their project Natick \cite{dunkingdc} which allows them to operate the data centre in an energy-efficient manner by leveraging heat-exchange techniques with outside water. This technique is commonly used in submarines. This experimental project shows that immersion cooling is viable in large scale computing systems with a group of servers sealed into large submarine cabinets. 

\par
Some other techniques have also explored but rarely used in large scale settings, such as Dielectric fluid, where server components are coated with a non-conductive liquid. The heat is removed from the system by circulating liquid into direct contact with hot components, then through cool heat exchangers. Such methods are not widely adopted yet in practice. The common issue with rack-level liquid cooling is a lack of standardisation and specifications among multi-vendors. However, due to its energy efficiency compared to air cooling, it is expected that liquid cooling would become mainstream in future data centre cooling systems. 

\section{Future Research Directions} \label{Sec_Future}
Cloud computing can be further improved by addressing several key issues that require detailed investigation and solutions.  This section gives some insights into these challenges for future work in this area.
\subsection{Moving from "time-to-solution" to "Kw-to-solution"} 
The current software development paradigms, platforms, and algorithms focus on improving applications' execution speed, neglecting its energy footprints. Hence, a paradigm shift is required to move from  ``time-to-solution'' to ``Kw-to-Solution'' in software development and deployments. New tools and programming constructs are needed to facilitate software developers to analyse the energy footprints of application logic so that developers can optimise software applications to minimise energy and improve execution speed. 
\subsection{Standardisation and Tools for AI-centric  RMS}
One of the important obstacles in adopting AI or ML solutions in data centre RMS is the lack of standardisation and tools. ML solutions need a good amount of data. Currently, distributed systems, including Cloud systems, produce vast amounts of data belonging to different computing layers. Standard methods and semantics are needed to collect, monitor, and interpret these data to adopt AI-centric models faster. Moreover, software tools and libraries need to be built specifically to resource management systems, which will easily integrate policies into existing systems.

\subsection{Cross-Layer Coordination Methods in Cloud Computing Stack} 
The total energy efficiency is achieved when resources are managed efficiently across different computing layers from on-chip microprocessor, data centre level platforms. Current approaches are limited to the individual computing layer due to a lack of coordination and heterogeneity among different computing layer. New interfaces and APIs can be built that easily facilitates and allows configuration across different computing layers.
\subsection{Resource Management in Emerging Cloud Execution Models}
As Cloud computing is evolving, it is moving from partially managed services to fully-managed services through application execution models such as Serverless computing. Serverless computing allows an application to be built based on multiple stateless microservices. Cloud service providers manage these microservices or stateless functions lifecycle completely with an assurance of autonomic scalability. It brings new challenges in pricing and the management of thousands of stateless application services.  This requires new resource management approaches in these fine-grained, network-accessed hardware resources shared by different containerised applications belonging to other users. 

\subsection{Holistic Resource Management}
Cloud data centres host closely interconnected systems, including computing, networking, storage and cooling systems. All these systems are closely interconnected and play an essential role in reliable service delivery. The resource management system should identify the dependencies and manage the resources holistically to achieve higher energy efficiency. It requires the development of new algorithms and platforms that configure parameters across different systems managing tradeoffs. 

\subsection{Efficiency Across Multi-tier Computing Platforms} 
The emergence of multi-tier (distributed computations from the network edge to remote clouds)  computing paradigms such as Edge/Fog computing to support IoT applications has created new resource and application management challenges. Applications in such environments require low latency response, which requires Cloud services to move from centralised remote locations to the network's edge. Such environments are highly heterogeneous than remote Clouds and are powered through battery or limited energy sources. Hence, application and resource management under these resource-constrained environments is challenging, requiring new solutions and approaches. 
\subsection{Decarbonising Cloud Computing}
Cloud data centres contribute significant CO\textsubscript{2} emissions due to their heavy reliance on brown or fossil fuel-based energy sources. Many service providers are procuring renewable energy to decarbonise Cloud systems. However, intermittent availability has hindered the adoption of renewable energy sources. New solutions shall explore addressing energy storage infrastructures and workload management in uncertain energy availability. Moreover, policymakers need to enforce new regulations for  Cloud service providers to adopt greener energy sources to power their data centres.

\subsection{Privacy-aware Resource Management}
The increasing security threats to internet services have brought new challenges in managing digital platforms. The new regulations, such as the General Data Protection Regulation (GDPR), require the data to be stored within the data source's geographic jurisdiction. This necessitates resource management solutions to be privacy-aware, requiring distributed storage and multi-part computation or computation over partial data. Hence,  resource management platforms should be built considering such privacy and security requirements of applications.

\section{Conclusions} \label{chap2-Sec:summary}
Cloud computing platforms are massively complex, large scale, and heterogeneous, enabling the development of highly connected resource-intensive business, scientific, and personal applications. Data centres have become a backbone infrastructure of the Cloud. Holistic energy and thermal management in such complex infrastructure have become a challenging task. The state-of-the-art rule-based or heuristics resource management solutions have become inadequate in modern Cloud data centres. The RMS policies need to deal with massive scale, heterogeneity, and varying workload requirements. Hence, we require data-driven AI approaches that derive key insights from the data, learn from the environments, and take resource management decisions accordingly. In this paper, we have discussed the challenges associated with adopting AI-centric solutions from the perspectives of energy and thermal management. We provided taxonomy for energy,  thermal and integrated resource management in Cloud data centres. Based on taxonomy, we have presented a detailed survey focusing on diverse resource management techniques.  Finally, we presented key future research directions.

\bibliographystyle{ACM-Reference-Format}
\bibliography{references}

\appendix

\end{document}